\documentclass[citeautoscript,showpacs,12pt,prb,aps]{revtex4}
\usepackage{bm,graphicx,amsmath,amssymb,color}
\newcommand{\avec}{\mathbf{a}}

\newcommand{\kvec}{\mathbf{k}}

\newcommand{\uvec}{\mathbf{u}}

\newcommand{\Uvec}{\mathbf{U}}

\newcommand{\vvec}{\mathbf{v}}
\newcommand{\Vvec}{\mathbf{V}}

\newcommand{\Fvec}{\mathbf{F}}

\newcommand{\Ovec}{\mathbf{\Omega}}
\newcommand{\Pvec}{\mathbf{\Pi}}
\newcommand{\xvec}{\mathbf{x}}

\newcommand{\Lmat}{\mathcal{L}}
\newcommand{\Mmat}{\mathcal{M}}
\newcommand{\Nmat}{\mathcal{N}}
\newcommand{\Pmat}{\mathcal{P}}
\newcommand{\Rmat}{\mathcal{R}}
\newcommand{\Smat}{\mathcal{S}}
\newcommand{\Tmat}{\mathcal{T}}
\newcommand{\curl}{\text{curl}}
\newcommand{\diver}{\text{div}}
\newcommand{\ez}{\mathbf{e}_z}
\newcommand{\lansa}{LANS$-\alpha$}

\newcommand{\laeba}{LAEB$-\alpha$}
\newcommand{\zerovec}{\mathbf{0}}
\newcommand{\half}{\frac{1}{2}}

\newcommand{\avecp}{\mathbf{a}_\perp}
\newcommand{\apll}{a_\parallel}
\newcommand{\kvecp}{\mathbf{k}_\perp}
\newcommand{\kpll}{k_\parallel}
\newcommand{\Lmatp}{\mathcal{S}_\perp}
\newcommand{\newt}[1]{\textcolor{blue}{#1}}

\begin{document}
\title{Elliptic instability in the Lagrangian-averaged 
Euler-Boussinesq-alpha equations}
\author{Bruce R. Fabijonas$^1$ and Darryl D. Holm$^{2,3}$}
\affiliation{%
$^1$Department of Mathematics, 
Southern Methodist University, 
Dallas, Texas 75275\\
$^2$Computer and Computational Sciences Division, 
Los Alamos National Laboratory,
Los Alamos, New Mexico 87545 \\
$^3$Mathematics Department, 
Imperial College, 
London SW7 2AZ, United Kingdom
}%
\date{\today}
\begin{abstract}
We examine the effects of turbulence on elliptic instability of
rotating stratified incompressible flows, in the context of the
Lagrangian-averaged Euler-Boussinesq-alpha, or \laeba, 
model of turbulence. We find that the \laeba~model
alters the instability in a variety of
ways for fixed Rossby number and Brunt-V\"ais\"al\"a frequency.  
First, it alters the location of the instability
domains in the $(\gamma,\cos\theta)-$parameter 
plane, where $\theta$ is the angle of incidence the Kelvin wave makes
with the axis of rotation and $\gamma$ is the eccentricity
of the elliptic flow, as well as the size of the associated Lyapunov
exponent.  Second, the model
shrinks the width of one instability band while simultaneously
increasing another.  Third, the model introduces bands of unstable
eccentric flows when the Kelvin wave is two-dimensional.  
We introduce two similarity variables--one is a
ratio of the Brunt-V\"ais\"al\"a frequency to the model parameter
$\Upsilon_0 = 1+\alpha^2\beta^2$, and
the other is the ratio of the adjusted inverse Rossby number to the
same model parameter.  Here, $\alpha$ is the turbulence correlation
length, and $\beta$ is the Kelvin wave number.
We show that by adjusting the Rossby number and
Brunt-V\"ais\"al\"a frequency so that the similarity variables remain
constant for a given value of $\Upsilon_0$, turbulence has little
effect on elliptic instability for 
small eccentricities $(|\gamma| \ll 1)$.  For moderate and large
eccentricities, 
however, we see drastic changes of the unstable
Arnold tongues due to the \laeba~model.
\newt{Additionally, we find that introducing anisotropy in the
  vertical component of the transported velocity field merely alters
  the definition of the model parameter $\Upsilon_0$, which
  effectively reduces the original parameter value.  When the
  similarity variables are viewed with the new definition, the results
  are similar to those for the isotropic case.}
\end{abstract}
\maketitle

\clearpage
\section{Introduction}
This paper is the third in a continuing investigation of the effects
of a turbulence closure model on the classic solution of elliptic
instability.  The first \cite{fab:holm:03} examined the effects of the
Lagrangian-averaged Navier-Stokes-alpha, or \lansa, turbulence closure
model in an
inertial frame.  The second \cite{fab:holm:04a} examined the effects
of this turbulence model in a rotating frame, as well as the effects of
a few related turbulence closure models.  In the present paper, we continue
the investigation by examining the effect of stratification with
rotation in the inviscid \lansa~model, also called 
the Lagrangian-averaged Euler-Boussinesq-alpha, or
\laeba, model \cite{holm:marsden:rat:98a,holm:99,holm:mar:rat:02}. 

\newt{%
Turbulence models are ubiquitous in fluid mechanics today.  They are
used with the hopes that they adequately describe the evolution of
real-world flows, governed by the Navier-Stokes equations.  
The question that arises, then, is how do
individual turbulence models affect classical solutions of the Navier-Stokes
equations?   A natural thought process is that a model which preserves
classical solutions is preferred over a model which drastically alters the
solutions.  However, imposing such a strict requirement on a turbulence
model is unrealistic.    
An alternative thought process asks how to change the
problem so that the effects of the model on classical solutions is
negated while using the low-resolution advantage  the
turbulence model possesses.  Our present investigation has two goals. 
First, we seek to understand how
the \laeba~turbulence model affects the classical solution known as elliptic
instability.   The model is formulated in a way that preserves the
existence of the solution, something which is not possible with every
turbulence model.  We find that the model alters the instability, which
leads to our second goal.  
Namely, we seek a reformulation of the problem so that the effects are
negated.  
Turbulence models are being used in applications such as ocean
circulation.  
Our motivation is to determine how turbulence closure models affect
instabilities {\it before} implementing the models in such
applications.}

The \laeba~model is a turbulence closure model which has two
velocities, a transport velocity and a transported velocity.  
The transport velocity $\uvec$ is smoother that the velocity $\vvec$
(or momentum) which it transports, given by the relation 
$\vvec = (1-\alpha^2\Delta)\uvec$ in the isotropic version. (An
anisotropic version will be given in Section 5.)    
This approach was first introduced
by Leray \cite{leray:34} as a regularization of the Navier-Stokes
equations.  
Here, the parameter $\alpha$ is interpreted as the turbulence
correlation length.  The idea is that in numerical simulation,
$\alpha$ is set to be the grid size, and that the dynamics which occur
at scales smaller than $\alpha$ are swept along to scales larger than
$\alpha$ by the \newt{transport velocity} $\uvec$.  

Elliptic instability is a mechanism by which two-dimensional swirling
flows become unstable to three-dimensional perturbations.  The
instability in the Euler equations was
investigated by Lord Kelvin \cite{kelvin} for the case of
circular streamlines and by Bayly \cite{bayly:86} for elliptic
streamlines.  Further work by others \cite{craik:crim:86, craik:88,
  craik:89, wal:90, miya:fuku:92, kers:93,  miya:93, leb:03} examined
the effects of rotation, stratification, and magnetic fields on the
instability.  \newt{We refer the reader to a recent
  review\cite{kers:02} for details about the history of elliptic
instability.}  It is a member of the Craik-Criminale, or CC, class of
exact solutions to the Navier-Stokes equations \cite{craik:crim:86}.
The velocity field in this class of solutions is the sum of a flow
linear in the spatial coordinates and a Kelvin wave. The present
authors have shown that this class of solutions are valid in a family
of turbulence closure models \cite{fab:holm:04a}.  This investigation
is continued here.  
 
The paper is organized as follows.  In Section~2, we state the
\laeba~equations and  discuss the
CC class of solutions for these equations.  We
examine a specific member of this class in Section~3, namely elliptic
instability.  We combine
analysis and numerical simulation to
give a full description of elliptic instability in the
isotropic \laeba~model in Section~4.    
The analysis yields two similarity variables which fully describe the
behavior of the instability for elliptic flows of small eccentricity.
\newt{We briefly discuss the effects of the anisotropic version of the
  \laeba~model in Section~5.  We show that the effects of anisotropy
  are similar to the isotropic version, once the similarity variables
  of Section~4 are redefined.}  
Finally, Section~6 concludes the paper with a summary of our findings,
including a conjecture as to why the model alters the unstable Arnold
tongues for moderate and large eccentricities.  

\section{CC solutions in \laeba~equations}
The isotropic version of the \laeba~equations
\cite{holm:marsden:rat:98a,holm:99,holm:mar:rat:02} are
\begin{align}
&\partial_t \vvec + \uvec\cdot\nabla\vvec + (\nabla\uvec)^T\cdot\vvec
 + 2\Ovec\times\uvec + bg\ez
+ \nabla \left ( p - \half|\uvec|^2 - \frac{\alpha^2}{2}|\nabla\uvec|^2
    \right )  = \Fvec, 
\label{eq:eb1}\\
&\partial_t b + \uvec\cdot\nabla b = 0,  \label{eq:eb2}\\
&\diver\,\uvec = 0 . \label{eq:incomp1}
\end{align}
The variables are as follows:
\begin{itemize}
\item $\vvec = (1-\alpha^2\Delta)\uvec$ is the fluid velocity field,
\item $\uvec$ is the mean transport fluid velocity field,
\item $p$ is the pressure,
\item $b=b(z)$ is the dimensionless buoyancy term which reflects the
  deviation from mean density [that is, we have taken
$\rho = \rho_0(1+b)$],
\item $\Ovec = \Omega\ez$ is the Coriolis force, 
\item $g$ is the gravitational constant,
\item $\Fvec$ is the contribution of all external body forces,
\item in index notation, $(\nabla\uvec)_{ij} = \partial_j u_i = u_{i,j}$.
\end{itemize}
The equations reduce to the classic Euler-Boussinesq (EB) equations
upon setting $\alpha= 0$.  
The equations admit solutions $(\uvec,p,b) = (\Uvec_0,P_0,b_0)$ given
by  
$\Uvec_0 = \Smat(t)\cdot\xvec + \bar\Uvec(t)$, 
$P_0 = \xvec^T\cdot\Tmat(t)\cdot\xvec + P(t)$, and
$b_0(z) = s(h-z)$, provided
\begin{eqnarray}
&d_t \Smat_{ij} + \Smat_{ir}\Smat_{rj} + 2\Omega\epsilon_{i3p}\Smat_{pj} 
-sg\delta_{3i}\delta_{3j} =  \Mmat_{ij}, \\
&\Smat_{ii} = 0,
\end{eqnarray}
where we sum over repeated indices.  
Here, $\Vvec_0 = (1-\alpha^2\Delta)\Uvec_0 = \Uvec_0$, and the
(symmetric) Hessian $\Mmat$ is defined by $\Mmat_{ij} =
-\partial_i\partial_j\mathbb{P}$, where 
\begin{eqnarray*}
\mathbb{P} = p -\int \Fvec\cdot\, d\xvec + (d_t\bar U_q -
\Smat_{qr}\bar U_r +shg\delta_{3q})x_q \, .
\end{eqnarray*}
The Brunt-V\"ais\"al\"a frequency, defined for the 
EB equations as $N^2 =  -(g/\rho_0)(d\rho/dz)$, is $N^2=sg$ with $s>0$.  

One can construct a new exact solution of the \laeba~equations of the form
\begin{eqnarray}\label{eq:CCa}
\uvec = \Uvec_0 + \Uvec_1, \quad
b = b_0 + b_1, \quad
p = P_0 + P_1 . 
\end{eqnarray}
Consequently, $\vvec = \Vvec_0 + \Vvec_1$, where $\Vvec_1 =
(1-\alpha^2\Delta)\Uvec_1$.  
The generic equations of motion for $\Vvec_1$, $b_1$, and $P_1$ are
\begin{widetext}
\begin{align}
\begin{split}
&\partial_t\Vvec_1 + \Uvec_0\cdot\nabla\Vvec_1 + \Uvec_1\cdot\nabla\Vvec_0
  + \Uvec_1\cdot\nabla\Vvec_1  
  + (\nabla\Uvec_0)^T\cdot\Vvec_1 + (\nabla\Uvec_1)^T\cdot\Vvec_0
\\ &\hspace{2em}  
  + (\nabla\Uvec_1)^T\cdot\Vvec_1  
  + 2\Ovec\times\Uvec_1 + b_1g\ez -
  (\nabla\Uvec_0)^T\cdot\Uvec_1 - (\nabla\Uvec_1)^T\cdot\Uvec_0  
\\ &\hspace{2em}  
   + \nabla \left ( P_1 - \half|\Uvec_1|^2 -
  \frac{\alpha^2}{2}|\nabla\Uvec_1|^2 -
  \alpha^2(\nabla\Uvec_0)_{ij}(\nabla\Uvec_1)_{ij} \right ) = \zerovec ,
\end{split}\label{eq:dista}\\
& \partial_t b_1 + \Uvec_0\cdot\nabla b_1 + \Uvec_1\cdot\nabla b_0 +
  \Uvec_1 \cdot\nabla b_1 = 0, \label{eq:distc} \\
&\diver\,\Uvec_1 = 0. \label{eq:incomp}
\end{align}
%
Equations~\eqref{eq:dista} and \eqref{eq:distc} contain two nonlinear
terms.  
We choose the disturbance for the buoyancy and transport
velocity field $\uvec$ 
in the form of a traveling Kelvin wave:
\begin{gather}
\Uvec_1 = \mu\avec(t) e^{i\beta\psi},
\quad b_1 = s\mu \hat b(t)e^{i\beta\psi}, 
\label{eq:CCb}  \qquad 
P_1 =  \mu \hat p_1(t)e^{i\beta\psi}
   +\mu^{2}\hat p_2(t)e^{2i\beta\psi}, 
\end{gather}
where $\psi(\xvec,t) = \kvec^T(t)\cdot\xvec+\delta(t)$, 
$\beta$ is the wave number chosen so that $|\kvec(0)| = 1$, and 
$\mu$ is a scaling parameter so that $|\avec(0)|=1$.  
Consequently, the transported velocity field has the form
$\Vvec_1 = \mu\Upsilon(t)\avec(t) e^{i\beta\psi}$, 
where  
\begin{eqnarray}\label{eq:upsilondef}
\Upsilon(t) = 1+\alpha^2\beta^2|\kvec(t)|^2.
\end{eqnarray}
We refer to
solutions of the form \eqref{eq:CCa} with \eqref{eq:CCb} as
Craik-Criminale, or CC,
solutions \cite{craik:crim:86}.  A CC solution can be viewed as the
sum of a `base flow' $\Uvec_0$ 
plus a `disturbance' $\Uvec_1$ in the form of a traveling Kelvin wave.  
The incompressibility condition \eqref{eq:incomp} yields
transversality of the wave
\begin{eqnarray}
\kvec^T\cdot\avec = 0. \label{eq:incompcond}
\end{eqnarray}
Upon inserting \eqref{eq:CCb} into \eqref{eq:dista} and
\eqref{eq:distc}, 
the terms in \eqref{eq:dista} and \eqref{eq:distc} nonlinear 
in the disturbance identically vanish due to wave transversality,
implied by \eqref{eq:incompcond}. 
The remaining terms are linear and constant in $\xvec$. The CC
solution ansatz requires that the coefficients of these terms vanish
separately, thereby yielding the following system of equations for the
wave vector $\kvec$, wave amplitude $\avec$, and pressure harmonics
$\hat p_{1,2}$:
\begin{align}
&d_t\kvec^T\cdot\xvec + \kvec^T\cdot\Smat\cdot\xvec = 0, \label{eq:press3}\\
\begin{split}
&
d_t(\Upsilon\avec) 
   + i\beta\Upsilon\avec(d_t\delta + \kvec^T\cdot\bar\Uvec) 
   +\Smat^T\cdot(\Upsilon\avec) 
   + \Pvec\times\avec - (i\beta
   \hat p_1-\alpha^2\beta^2\kvec^T\cdot\Smat\cdot\avec)\kvec
 \\ &\hspace{1em} 
  + N^2\hat b\ez = \zerovec,
\end{split}
 \label{eq:aevolg} \\
&sd_t \hat b + is\beta \hat b(d_t\delta + \kvec^T\cdot\bar\Uvec) 
  + (\nabla b_0)\cdot\avec = 0, \label{eq:cevolg}\\
&\hat p_2 + (\Upsilon-1)|\avec|^2 = 0. \label{eq:p2}
\end{align}
\end{widetext}
Here, $\Pvec = 2\Ovec + \curl~\Uvec_0$ is the total vorticity of the
rotating coordinate system and the base flow.
Without loss of generality, we set  
$d_t\delta + \kvec^T\cdot\bar\Uvec = 0$.  We take the gradient of
\eqref{eq:press3} and obtain the following system of equations:
\begin{align}
& d_t\kvec + \Smat^T\cdot\kvec = \zerovec,\label{eq:kevolg}\\
&
d_t(\Upsilon\avec) 
   +\Smat^T\cdot(\Upsilon\avec) 
   + \Pvec\times\avec - \tilde P\kvec 
  + N^2\hat b\ez = \zerovec,
 \label{eq:aevolg1} \\
&d_t \hat b  
  - \avec\cdot\ez = 0, \label{eq:cevolg1}
\end{align}
subject to \eqref{eq:incompcond}.  The term $\tilde P$
represents the coefficient of $\kvec$ in \eqref{eq:aevolg}.   
At this point, one may assume that $\avec(t)$, $\hat b(t)$,
$i\hat p_1(t)$, and $\hat p_2(t)$ are
real-valued functions. 

The operator $(d_t + \Smat^T\cdot)$ in \eqref{eq:kevolg} and
\eqref{eq:aevolg1} is the total derivative
of a vector field in a Galilean frame moving with $\Uvec_0$.  Thus, 
these equations state that the wave vector $\kvec$ is frozen 
in the fluid, and the scaled amplitude $\Upsilon\avec$ in the Galilean frame
moving with $\Uvec_0$ is affected by the total vorticity of the
system, by pressure, by the buoyancy, and by the base flow.
Once the CC solutions are established, introduction of the factors of
$\Upsilon$ in equations \eqref{eq:p2} and \eqref{eq:aevolg1} reveals  the key
differences 
between the elliptic instability analyses for the EB equations and
the mean turbulence equations of the \laeba~model. 
 
As is common when working with incompressible fluids, 
the pressure term can be removed
from the problem by taking the dot 
product of \eqref{eq:aevolg1} with $\kvec$ and using the fact that 
$\kvec^T\cdot\avec$ is an integral of motion: 
\begin{eqnarray*}
\tilde P = \frac{ \kvec^T\cdot  ( 
   \Upsilon(\Smat+\Smat^T)\cdot\avec 
  +\Pvec\times\avec + N^2\hat b\ez  )}{|\kvec|^2}.  
\end{eqnarray*}
Once \eqref{eq:kevolg} 
is solved for $\kvec$, the equations for $\avec$ and $\hat b$ can be written
compactly as
\begin{eqnarray}\label{eq:system}
\frac{d}{dt}\begin{pmatrix} \avec  \\ \hat b \end{pmatrix}
  = \Nmat(t;\ldots ) \begin{pmatrix} \avec \\ \hat b \end{pmatrix} .
\end{eqnarray}
Here, $\Nmat$ is a $4\times 4$ matrix whose coefficients contain
elements from $\Smat$ and $\kvec$, and the dots indicate parameter
dependence.  

\paragraph*{Multi-harmonic disturbances.}  
Finally, we point out that for the EB equations (corresponding to
$\alpha = 0$), the perturbation can be over a sum of harmonics of
$e^{i\beta\psi}$.  In this case, different harmonics do not interact.  
This phenomenon changes slightly 
for the \laeba~equations.  For example, if we
were to consider $\Uvec_1 = \mu\avec^{(1)}\exp(i\beta\psi) +
  \mu^2\avec^{(2)}\exp(2i\beta\psi)$ and similarly for $\Vvec_1$ and
$b_1$, then the correct form for the pressure perturbation $P_1$
would carry the first {\it four} harmonics of $\exp(i\beta\psi)$.  The
pressure coefficients $\hat p_i(t)$ 
of the first, third and four harmonics would be purely
real.  However, the coefficient of the second harmonic would be
complex-valued such that the real part balances a contribution
from $\avec^{(1)}$ and the imaginary part from $\avec^{(2)}$.  
Thus, different harmonics do interact, but the interaction is absorbed
by the pressure.  The remainder of the paper considers only
single-harmonic disturbances. 

\section{Elliptic instability}
The classic problem of elliptic instability resides among the CC
solutions. The base flow for elliptic instability is 
\begin{eqnarray}\label{eq:ellins}
\Smat = \bar\omega\begin{pmatrix} 0 & -1+\gamma & 0 \\
               1 + \gamma & 0 & 0 \\
                        0 & 0 & 0 \end{pmatrix}, \qquad \bar\Uvec = \zerovec.
\end{eqnarray} 
This base flow is a rigidly rotating column of fluid whose
streamlines are ellipses with eccentriticy $\gamma$ and vorticity
$2\bar\omega\ez$.  We rescale time by $\bar\omega$ so that
the Coriolis parameter $\Omega$ is replaced by $\Omega' =
\Omega/\bar\omega$, which we interpret as an inverse Rossby number
\cite{miya:fuku:92,fab:holm:04a}.   We now suppress the prime notation.
The extreme  
values of the solution in \eqref{eq:ellins} 
represent pure shear at $\gamma = \pm 1$ (along different axes) 
and rigid body rotation at $\gamma = 0$. 
The solution of \eqref{eq:kevolg} is 
\begin{eqnarray}
\kvec = \begin{pmatrix} \sin\theta\cos \hat t   \\
               \kappa\sin\theta\sin\hat t\\
                \cos\theta \end{pmatrix},
\end{eqnarray}
where $\hat t = t\sqrt{1-\gamma^2}$, $\kappa^2 = (1-\gamma)/(1+\gamma)$,
and $\theta$ is the polar
angle the wave vector $\kvec$ makes with the axis of rotation $\ez$.
We note that a more arbitrary initial orientation of $\kvec$ is
equivilaent to a shift in the starting time $t\to t-t_0$.  

In general, the solution of \eqref{eq:system} must be simulated
numerically.  Since the coefficient matrix $\Nmat$ in
\eqref{eq:system} is periodic with 
period $\tau = 2\pi/\sqrt{1-\gamma^2}$, it follows that 
\eqref{eq:system} can be rewritten as a pair of  
Schr\"odinger equations with periodic potentials, shown in
the appendix.  
Thus, we analyze the system using Floquet theory
\cite{yaku:star:76}.
One needs to determine the eigenvalues
$\rho_i$ of the monodromy matrix 
$\Pmat(\tau)$, where $d_t\Pmat = \Nmat\cdot\Pmat$ with
the initial condition $\Pmat(0) = \mathcal{I}_4 =
\text{diag}\{1,1,1,1\}$.  
The Lyapunov growth
rates, which exist only if $\max_i|\rho_i| > 1$, are given by $\sigma =
\ln(\max_i|\Re(\rho_i)|)/\tau$.  

\section{Isotropic \laeba~model}
We begin by examining the isotropic model presented in Section II.
The classical results for the EB 
equations were obtained by Miyazaki and
Fukumoto\cite{miya:fuku:92} and Miyazaki\cite{miya:93}.
Rather than review the classical results separately, we will study the
solution in the isotropic \laeba~model directly, regaining the
classical results upon setting   
$\alpha = 0$, which is equivalent to setting $\Upsilon(t) \equiv 1$.  To
illustrate the effects of the model, we plot the classical results
next to those generated by the model.  The {\it pi\`ece de
  r\'esistance} of the present paper is the introduction of the
similarity variables 
\begin{equation}\label{eq:newdefs}
\chi = \frac{N^2}{\Upsilon_0}, \qquad \zeta = \frac{2|\Omega+1|}{\Upsilon_0},
\end{equation}
where 
\begin{equation}\label{eq:upzero}
\Upsilon_0 = 1+\alpha^2\beta^2.
\end{equation}
Since $\alpha$ has dimensions of length, it follows that $\Upsilon_0$
is a dimensionless parameter.  

\subsection{Circular streamlines--exact solutions}
The circular case $\gamma=0$ can be analyzed analytically.  Here, $|\kvec(t)| =
1$, and any solution to \eqref{eq:system} can
be written as  
\begin{widetext}
\begin{eqnarray} \label{eq:gen}
\begin{pmatrix} \avec  \\ \hat b \end{pmatrix} = 
  c_1 \begin{pmatrix} \avec_{1}  \\ \hat b_{1} \end{pmatrix} +
  c_2 \begin{pmatrix} \avec_{2}  \\ \hat b_{2} \end{pmatrix} +
  c_3 \begin{pmatrix} \avec_{3}  \\ \hat b_{3} \end{pmatrix} +
  c_4 \begin{pmatrix} \avec_{4}  \\ \hat b_{4} \end{pmatrix},
\end{eqnarray}
where
\begin{alignat*}{4}
\avec_{1} &= \sin(\omega t+\phi) \kvec_{\perp 1}  
     + \frac{1}{\sqrt{1+q^2\chi}}\cos(\omega t+\phi) \kvec_{\perp 2}, \quad
 &&\hat b_{1} = \frac{q}{\sqrt{1+q^2\chi}}\cos(\omega t+\phi) &\\
\avec_{2} &= -\cos(\omega t+\phi) \kvec_{\perp 1}  
     + \frac{1}{\sqrt{1+q^2\chi}}\sin(\omega t+\phi) \kvec_{\perp 2}, \quad
 &&\hat b_2 = \frac{q}{\sqrt{1+q^2\chi}}\sin(\omega t+\phi) &\\
\avec_{3} &= \begin{pmatrix} q^2\chi\cot\theta\cos t + q\chi t\sin t \\ 
          q^2\chi\cot\theta\sin t -q\chi t\cos t \\ 1 \end{pmatrix} , \quad
 &&\hat b_3 = t, &\\
\avec_{4} &= -q\chi\kvec_{\perp 2}, \quad
 &&\hat b_4 = 1. &
\end{alignat*}
\end{widetext}
Here, $\kvec_{\perp 1} = [\cos\theta\cos t, \cos\theta \sin t,
-\sin\theta]^T$, $\kvec_{\perp 2} = [-\sin t, \cos t, 0]^T$, 
$\omega = 2\zeta\cos\theta\sqrt{1+q^2\chi}$ and $q =
\tan\theta/\zeta$.  For $\alpha = 0$, 
these solutions are the exact version of the
$O(N^2)$ solutions of Kerswell\cite{kers:02}.  
Additionally, for $N^2 = 0$ ($\chi = 0$), we regain the
\lansa~solutions\cite{fab:holm:04a} for $\alpha > 0$ and the classic Euler 
solutions\cite{wal:90,kers:02} for $\alpha = 0$.  

To compute the monodromy matrix $\Pmat(2\pi)$, we use the solution in
\eqref{eq:gen} and determine the coefficients $c_i$ for each 
column of the monodromy matrix by the condition $\Pmat(0)=\mathcal{I}_4$.  The
resulting eigenvalues $\rho_i$ are $\rho_{1,2} = 1, \rho_{3,4} =
\exp(\pm 2i\omega\pi)$.
Since all of the eigenvalues satisfy $|\rho_i|= 1$, the circular
case ($\gamma = 0$) is stable.  Furthermore, the angle of critical
stability, that is, the parameter value at which instabilities will
set in as $\gamma$ increases from zero, is determined by the condition
$\rho_i = 1$, or equivalently, $2\omega\pi = \pm2n\pi$, where $n = 1, 2,
3, \ldots$.  This corresponds to 
\begin{eqnarray}\label{eq:critangle}
\cos\theta = \pm\sqrt{\frac{n^2-\chi}{\zeta^2-\chi}},\,\,\,n = 1,2, \ldots.
\end{eqnarray}
The first conclusion is that stratification and rotation
increases the angle of critical stability (decreases $\cos\theta$).  
The different values of $n$ indicate the primary, secondary, 
tertiary, etc., instability domains.  These different instability
domains, or fingers, have been extensively
studied\cite{miya:fuku:92,miya:93,fab:holm:04a}.   
The principal finger, $n=1$, is the main instability domain for
elliptic instability, and it is call the `subharmonic frequency.'  
The secondary finger, $n=2$, corresponds
to a resonance phenomenon with internal gravity waves, and thus is
called the `fundemental frequency.'  The remaining fingers,
$n=3,4,\ldots$, naturally are  referred to as `superharmonic
frequencies.'  In the absence of statification, 
only the subharmonic frequency ($n=1$) is physically interesting from 
the viewpoint of elliptic instability, 
although rotation increases the width of the other frequencies.
(These fingers are displayed in
Figs.~\ref{fig:baylyN2-EB}-\ref{fig:baylyN2z075-EB} below.) 
\newt{Since the expression for $\cos\theta$ is independent of $\Upsilon_0$
explicitly, the second conclusion we draw is that the
\laeba~model has no effect on the circular case, provided that
the values of $N^2$ and $\Omega$ change as $\alpha$ changes so that
$\chi$ and $\zeta$ remain fixed. }  

\subsection{Slightly elliptic streamlines--perturbation analyses}
The parameter values which yield critical stability for the circular
case will produce instability in the elliptic case for small
eccentricities.  We derive an analytical expression for
the leading order growth rate of this instability to $O(\gamma^2)$,
defined for the \laeba~model as  
\begin{eqnarray}
\sigma \equiv \frac{1}{2(\Upsilon_0|\avec|^2+N^2\hat b^2)}
  \frac{d}{dt}\Big (\Upsilon_0|\avec|^2+N^2\hat b^2\Big ).  
\end{eqnarray}
This is a natural extension of the definition given by Kerswell\cite{kers:02}
for $\Upsilon_0 = 1$, and differs slightly from that in
previous work on the \lansa~model\cite{fab:holm:03,fab:holm:04a}.  
By taking the dot product of \eqref{eq:aevolg1}
with $\avec$, multiplying \eqref{eq:cevolg1} by $N^2\hat b$, and
adding, we find that the following expression for $\sigma$:
\begin{eqnarray*}
\sigma = \frac{-\avec^T\cdot\Smat^T\cdot(\Upsilon_0\avec)}{2\Upsilon_0}
  = -\gamma a_1a_2
\end{eqnarray*}
valid for all $\gamma$, where $a_{1,2}$ are the first and second
components of $\avec$, respectively.  We
insert the circular solutions (that is, for $\gamma=0$) into the
above equation, say $\avec_1$ and $\hat b_1$, and average over a
period of the solutions.  Typically\cite{wal:90}, the average will
vanish except 
when $\omega\pm 1 = 0$, a condition which again yields
\eqref{eq:critangle}.  A maximum is attained when $\phi = \pm
\pi/4$, yielding
\begin{eqnarray}
\sigma_\text{max} 
= \Big (
\frac{9}{16}\times\frac{4(n^2-\chi)(\zeta+n)}{9\Upsilon_0n^2(\zeta^2-\chi)}
\Big )\gamma  + O(\gamma^2).  \label{eq:sigmax}
\end{eqnarray}
We see
that for base flows with small deviations from circular, that is, for
flows with  $|\gamma|\ll 1$, the growth rates are linear in the
eccentricity.  Of particular interest is the fact that
$\sigma_\text{max}$ is a function of all three parameter $\chi$,
$\zeta$, and $\Upsilon_0$, rather than just the first two.  
\newt{We conclude that although properly adjusting the
  Brunt-V\"ais\"al\"a frequency and Rossby number preserves the
  effects the model has on circular flows, the growth rate in the
  \laeba~model is a decreasing function of the model parameter $\Upsilon_0$.}
Numerical simulations below will show that for
moderate and large eccentricities, the effect is more dramatic.   
\newt{In particular, the subharmonic finger ($n=1$) has a
  maximum growth rate of $\sigma_\text{max} = \gamma/(2\Upsilon_0)$,
  which occurs 
  at $\zeta = 1$ ($\Omega = -1/2$).  For this parameter value, the
  critical angle is $\cos\theta =1$, that is, when the wave vector is
  parallel to the ellipse's axis of rotation.  This is referred as the
  {\it zero tilting vorticity}\cite{cambon:94,leb:03}.  We will expand
  on this case in a separate section.} 
  
\begin{figure}
\begin{center}
\includegraphics[angle=0,width=3in]{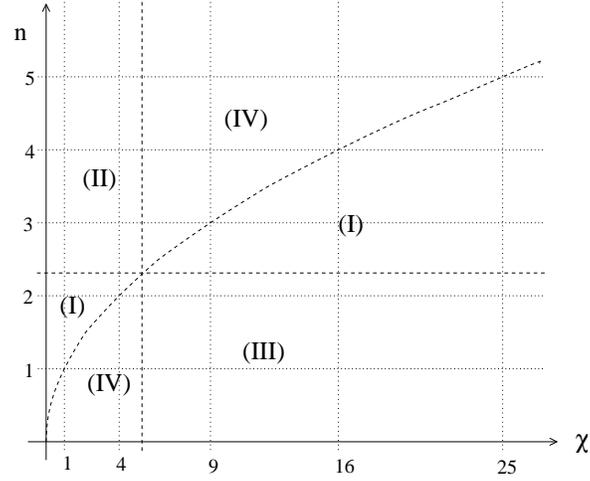}
\caption{The $(\chi,n)$ parameter plane for the representative fixed value
  $\zeta = \sqrt{5}$.  The  roman numerals indicate
  different regions described in the text.  Region (I) corresponds to
  real values of $\cos\theta$ which are less than unity.  Parameter
  values in this region will yield exponential growth of $\avec(t)$
  for small values of eccentricity $\gamma$.  
  Regions (II) and (III)
  correpsond to real values of $\cos\theta$ greater than unity, and
  region (IV) correpsonds to imaginary values of $\cos\theta$.  
  Parameter values which fall into these regions yield windows of
  eccentricity values for which the disturbance is always stable.  
  The curves separating different
  regions are the dashed curves given by the equations
  $n=\sqrt{\chi}$ (the parabola), $\chi =
  \zeta^2$ (the vertical line), and
  $n=\zeta$ (the horizontal line).  These curves separate the various
  types of elliptic instability behavior in the $(\chi,n)$ parameter
  plane into four regions, according to Eq.~\eqref{eq:critangle} for
  the critical angle. 
 \label{fig:N2n}}
\end{center}
\end{figure}

\begin{figure}
\begin{center}
\includegraphics[width=3in]{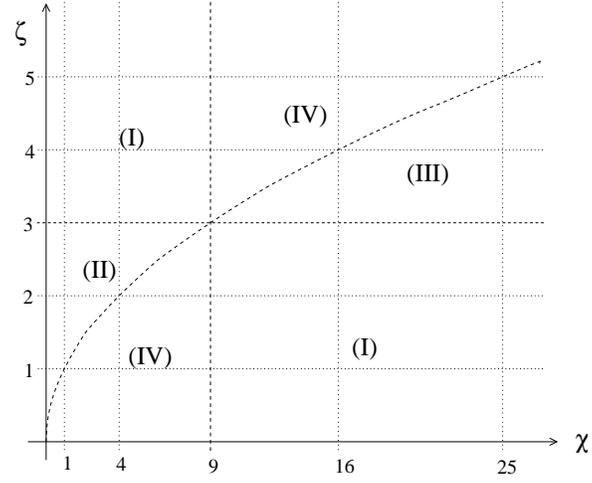}
\caption{The $(\chi,\zeta)$ parameter plane
  for the representative fixed value  $n=3$, 
with the different regions described in the text.
This figure describes the behavior of the tertiary finger for
  different values of $\zeta$ and $\chi$.   
The regions are separated by the dashed curves given by the parabola
$\chi=\zeta^2$, 
the vertical line $\chi=n^2$, and the horizontal
line $\zeta = n$.    \label{fig:Nomega}}
\end{center}
\end{figure}

The expression for the critical angle \eqref{eq:critangle}
separates the parameter space into four distinct
regions.  We describe each region by its characteristic behavior.  We
preserve the $\zeta$ and $\chi$ notation since the behavior will
remain unchanged under the redefinition of these two variables for the
\laeba~model in the next section.  
\begin{enumerate}
\item[(I)] Both the numerator and denominator of the
radicand have the same sign and the ratio is less than unity.
For parameter
values in this region, we have $\cos\theta < 1$.  Thus, 
slight perturbations in the
columnar vortex's eccentricity from circular will yield exponentially
growing amplitudes.  This is due to the fact that the fingers touch
the line $\gamma = 0$ at a point (a consequence of parametric
resonance and the fact that $\gamma = 0$ is stable).  For a given
finger with index $n$, we have that either $\cos\theta\to 0^+$ as $\chi\to
(n^2)^-$ and it vanishes for larger values of $\chi$ or 
$\cos\theta \to 1^-$ as $\chi\to\infty$, in which case it never vanishes.  
\item[(II)] Both the numerator and denominator of the
radicand are positive and the ratio is greater than unity.
Parameter values in this region correspond to $\cos\theta > 1$ and 
experience a window of nonzero eccentricities for which the Kelvin
waves are bounded (i.e. stability).  If we could extend the parameter
plane to include values of $\cos\theta$ larger than unity, then the
fingers which fall into region (II) would also experience
instability.  The unique property here is that $\cos\theta \to +\infty$
as $\chi\to(\zeta^2)^-$.
\item[(III)] Both the numerator and denominator of the
radicand are negative and the 
ratio is greater than unity.  
For these parameter values, the fingers whose index 
corresponds to $n<\zeta$ 
 have vanished and will not reappear.  They enter this region at $\chi =
 (\zeta^2)^+$ with $\cos\theta = +\infty$ and has the feature that
$\cos\theta\to 1^+$ as $\chi\to\infty$.
\item[(IV)] The numerator and denominator have different signs.  
 Parameter values which fall into
this region are particularly interesting.  For these values, 
 the instability domains align themselves vertically in the
$(\gamma,\cos\theta)-$plane rather than horizontally or diagonally
 towards $\cos\theta = 1$, $\gamma =0$. 
\end{enumerate}
Figures~\ref{fig:N2n} and \ref{fig:Nomega} shows the four regions in 
the $(\chi,n)$ and $(\chi,\zeta)$ parameter planes, respectively, and 
Figs.~\ref{fig:baylyN2-EB}-\ref{fig:baylyN2z075-EB} show numerical
simulations which exhibit the described behavior of the fingers. These
figures show domains of stable and unstable behavior as a function of
 buoyancy $\chi$ and  rotation $\zeta$ parameters for
elliptic instabilities of various orders;  subharmonic ($n=1$),
fundamental ($n=2$), and superharmonic ($n=3,4,\ldots$).  \newt{In
particular, subfigures (a)-(h) show the results for the EB equations
($\Upsilon_0 = 1$) and subfigures (i)-(p) show the effects of the
\laeba~model for the same values of the similarity variables $\zeta$
and $\chi$.  }  

Figure~\ref{fig:N2n} shows the different regions 
for a fixed representative value of $\zeta$.  
The horizontal and
vertical lines which separate the different regions intersect on the
parabola at the point
$(\zeta^2,\zeta)$.  
As $\zeta$ decreases, the intersection point slides
down the parabola towards the origin and the unstable fingers in
region (I) stabilize and enter region (II).  
The figure, as drawn, indicates that
the fingers corresponding to the subharmonic ($n=1$) and fundamental
($n=2$) frequencies  live in region (I)
for $n < \sqrt{\chi}$, in region (IV) for
$\sqrt{\chi} < n < \zeta^2$, and in
region (III) for $n > \zeta^2$. 
These fingers will be unstable in (I) and vanish in (IV) and (III).
In contrast, the
fingers for the superharmonic frequencies $n\geq 3$ live in region
(II) for $n < \zeta^2$, region (IV) for
$\zeta^2 < n < \sqrt{\chi}$, and finally
region (I) for $n > \sqrt{\chi}$. These fingers are aligned diagonally
in the $(\gamma,\cos\theta)$-plane for region (II) and vertically for
region (IV), and
then meet at the line $\gamma = 0$ as parameter values enter region
(I).    

We deduce two stability criteria from
Figs.~\ref{fig:N2n}-\ref{fig:Nomega} 
and \eqref{eq:critangle}.  First, for 
parameter values satisfying $\lfloor \zeta \rfloor < \sqrt{\chi} <
\lceil \zeta \rceil$, or in original variables, 
\begin{eqnarray}\label{eq:stabcrita}
\Big \lfloor \frac{2|\Omega+1|}{\Upsilon_0} \Big\rfloor 
  < \sqrt\frac{N^2}{\Upsilon_0} 
  < \Big\lceil \frac{2|\Omega+1|}{\Upsilon_0} \Big\rceil,
\end{eqnarray}
where $\lfloor x \rfloor$ and $\lceil x \rceil$ the floor and ceiling
of $x$, respectively, the flow is always stable.  Here, floor and
ceiling refer to rounding the the nearest integer less than and
greater than the current value, respectively. 
Essentially, \eqref{eq:stabcrita} follows from the fact that none of
the fingers 
live in region (I) for these parameter values. 
This criterion is analogous
to the nonresonance stability criterion $1/2 \leq
N/(2\Omega) \leq 2$  for
forced turbulence in three-dimensional rotating, stably stratified
flow in the Boussinesq approximation\cite{smith:wal:02}.  
(Note that their equation for the
buoyancy differs from ours.)  For $\zeta = \sqrt{5}$
in Fig.~\ref{fig:N2n}, this corresponds to $2 < \sqrt{\chi} < 3$.  
The second stability criterion is that  
all of the fingers will live in region (II) when
\begin{eqnarray}\label{eq:stabcritb}
-\frac{\Upsilon_0}{2} < \Omega+1 < \frac{\Upsilon_0}{2} \,\, \text{for}
 \,\, N^2 < \Upsilon_0,
\end{eqnarray}
Within this window, 
the flow is stable for slight perturbations in $\gamma$.  
This means counter rotation stabilizes the flow, when the
Brunt-V\"ais\"al\"a 
frequency is sufficiently low.  \newt{Contrast this with
  Leblanc's\cite{leb:03} stability criterion for the EB equations
  ($\Upsilon_0 = 1$), which says that for
  $\Omega = -1$, the flow is stable for parameter values satisfying
  $N^2\leq 1-\gamma^2$.  (In \eqref{eq:stabcritb}, $\gamma = 0$.) }

\subsection{Two-dimensional perturbations}
\subsubsection{$\cos\theta = 1$}
The case $\cos\theta = 1$, that is when the wave vector of the Kelvin
wave is parallel to the axis of rotation,  
also can be analyzed analytically.  \newt{As mentioned before, this is
referred to as the zero tilting vorticity because the vorticity of the
ellipse-wave system is merely stretched rather than tilted.  This follows
from the fact that $\kvec = [ 0, 0, 1]^T$, which, because of
transversality \eqref{eq:incompcond}, forces 
 $\avec = [ a_1, a_2, 0]$.  The resulting equations yield $\hat b(t) =
\text{const.}$, and }
\begin{align}
\dot a_1 + (1-\zeta+\gamma)a_2 &= 0, \nonumber \\
\dot a_2 - (1-\zeta-\gamma)a_1 &= 0.
\end{align}
It follows that no exponential growth of the solution of
\eqref{eq:system} will occur for
parameter values satisfying
\begin{equation}\label{eq:stabcrit3}
(1-\zeta)^2 - \gamma^2 \geq 0.
\end{equation}

\subsubsection{$\cos\theta = 0$}
\begin{figure}
\begin{center}
\includegraphics[width=3.4in]{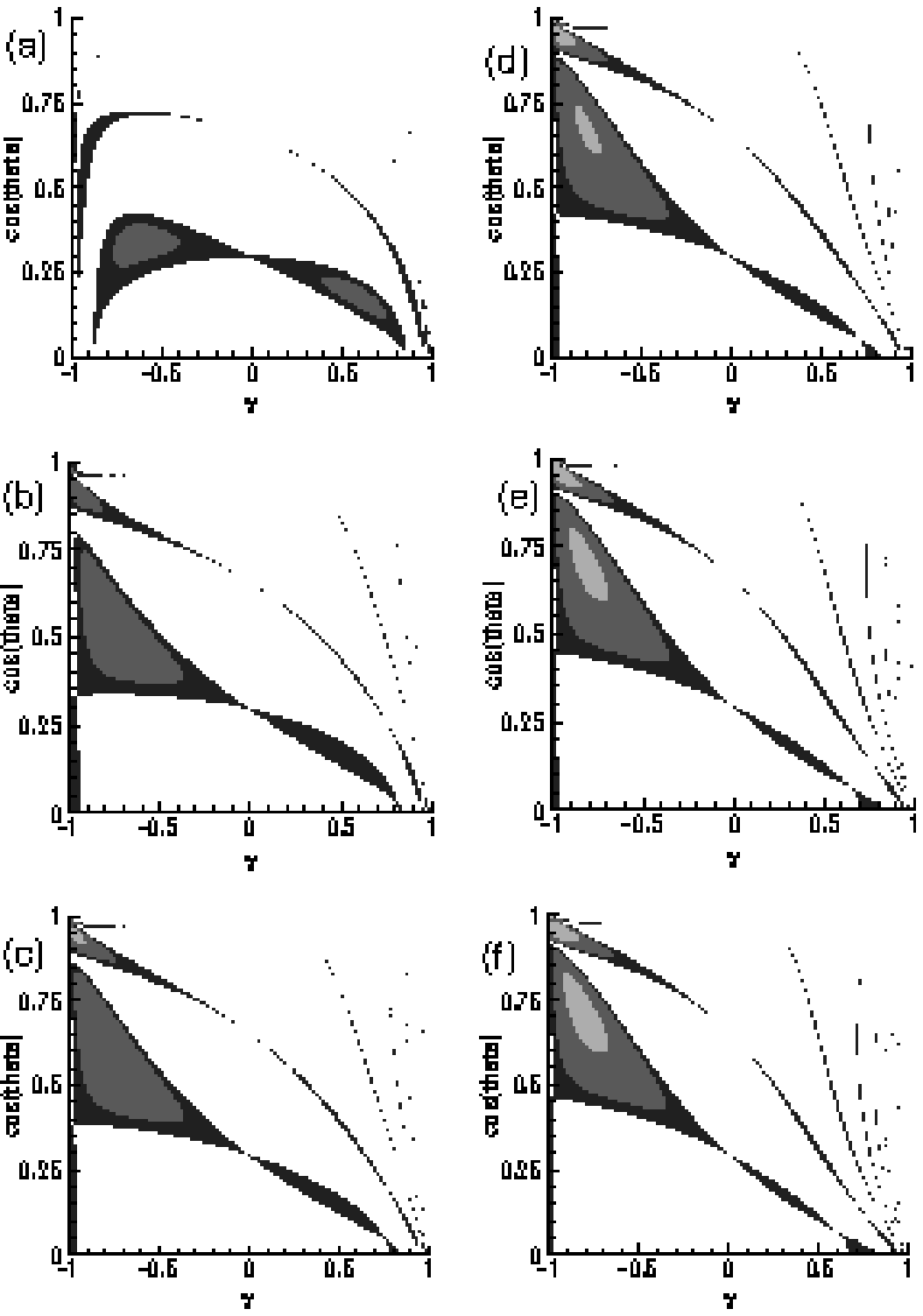}
\caption{Instability domain of $\avec(t)$ for 
   rotation parameter $\zeta = 3$, buoyancy parameter $\chi = 1/4$,
   and various  model parameter $\Upsilon_0$:
   (a)~1 (the EB equations), (b)~3/2, (c)~2,
   (d)~11/4, (e)~4, and (f)~6. \newt{The behavior of the
   instability 
   regions is essentially the same in all figures for $|\gamma| \ll
   1$.  }
\label{fig:baylyalpha1}
}
\end{center}
\end{figure}

\begin{figure}
\begin{center}
\includegraphics[width=3.4in]{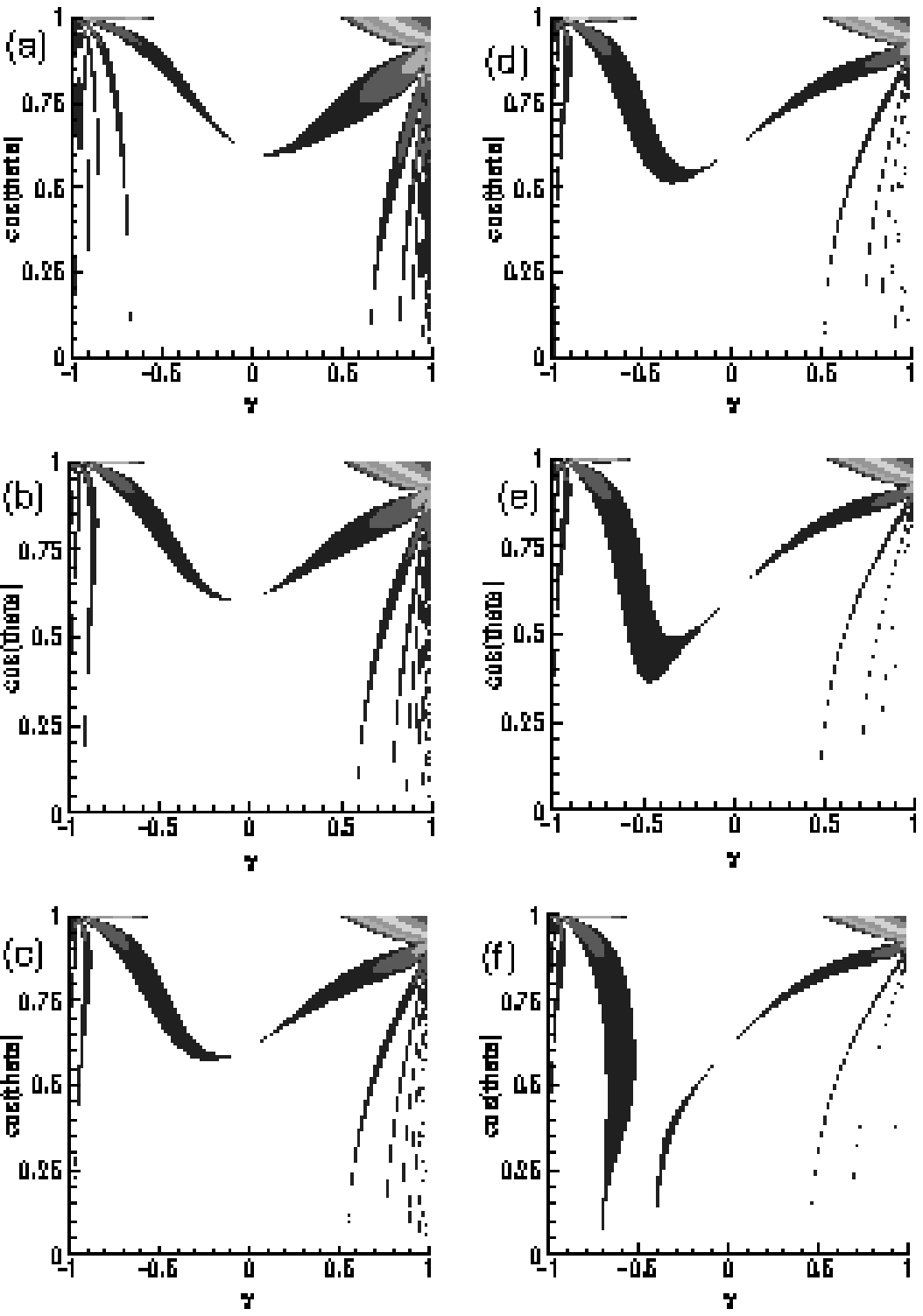}
\caption{Instability domain of $\avec(t)$ for 
   rotation parameter $\zeta = 3/2$, buoyancy parameter $\chi = 5$,
   and various  model parameter $\Upsilon_0$:
   (a)~1 (the EB equations), (b)~3/2, (c)~2,
   (d)~11/4, (e)~4, and (f)~6.  \newt{Again, the behavior of the instability
   regions is essentially the same in all figures for $|\gamma| \ll
   1$.  }
\label{fig:baylyalpha2}
}
\end{center}
\end{figure}

\begin{figure}
\begin{center}
\includegraphics[width=3.4in]{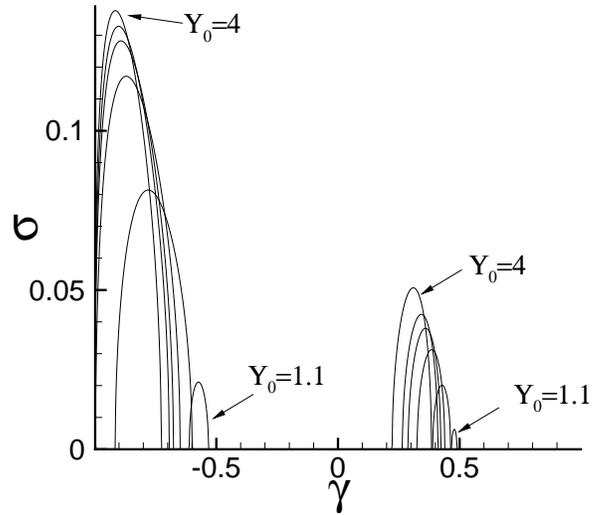}
\caption{The growth rate for the case $\cos\theta = 0$ for $\chi =
  3/4$ and $\zeta = 3$, and for values of $\Upsilon_0$ in the range $1
  \leq \Upsilon_0 \leq 4$.  For $\Upsilon_0 = 1$, the Lyapunov
  exponent is identically zero.  The exponent is immediately non-zero
  for $\Upsilon_0 > 1$, and shifts to the left as the parameter
  increases.  \newt{For larger values of $\Upsilon_0$, the growth rates
  approach an asymptotic limit, which is slightly to the left of the
  curve corresponding to $\Upsilon_0 = 4$ shown in the figure.  }
\label{fig:sigma_cth0} }
\end{center}
\end{figure}

We perform a separate
investigation for the case $\cos\theta = 0$, that is, when the wave
vector lies in the plane of the flow.  As was the case for $\cos\theta
= 1$, the equations
decouple such that the equations for $a_{1,2}$ are independent of
$a_3$ and $\hat b$, and vice-versa.  
The equations for $a_{1,2}$ are
exactly those for the \lansa~model, and 
the equations for $a_3$ and $\hat b$
can be rewritten as a single Schr\"odinger equation: 
\begin{eqnarray}\label{eq:schrodcth0}
\ddot v + \frac{N^2}{\Upsilon} v = 0,
\end{eqnarray}
where $v = \Upsilon a_3$ and overhead dot denotes a time derivative.  
For the EB case, $\Upsilon(t) = 1$, and the solutions are
purely periodic and bounded, i.e., stable.   
In fact, this case is critically stable in the sense
that any slight perturbation of the wave vector in the third dimension
($\cos\theta > 0$)
could result in Kelvin waves with exponentially growing amplitudes
\cite{miya:fuku:92,miya:93}.  \newt{This corresponds to the fingers touching
the $\gamma-$axis in the shape of a cusp, seen in subfigures (a)-(h) in
Figs.~\ref{fig:baylyN2-EB}-\ref{fig:baylyN2z075-EB},  
Fig.~\ref{fig:baylyalpha1}a, and
Fig.~\ref{fig:baylyalpha2}a, as predicted by Floquet theory.}
For the \laeba~model, \eqref{eq:schrodcth0} is once again a
Floquet problem.   
Comparison of numerical simulations of the full $4\times
4$ system in \eqref{eq:system} and that in \eqref{eq:schrodcth0} yield exactly
the same growth rates over the same parameter ranges.  We conclude,
then, that buoyancy coupled with the model parameter $\Upsilon_0$ is
the cause of this instability.   
This is illustrated in subfigures (i)-(p) of
Figs.~\ref{fig:baylyN2-EB}-\ref{fig:baylyN2z075-EB}, and in 
Fig.~\ref{fig:sigma_cth0}.  Numerical simulations suggest that these
two-dimensional instabilities exist only for $\chi < 2$, even for
extra-ordinarily large values of $\Upsilon_0$. 

\subsection{Elliptic streamlines--numerical simulations}
\begin{figure*}
\begin{center}
\includegraphics[width=6.8in]{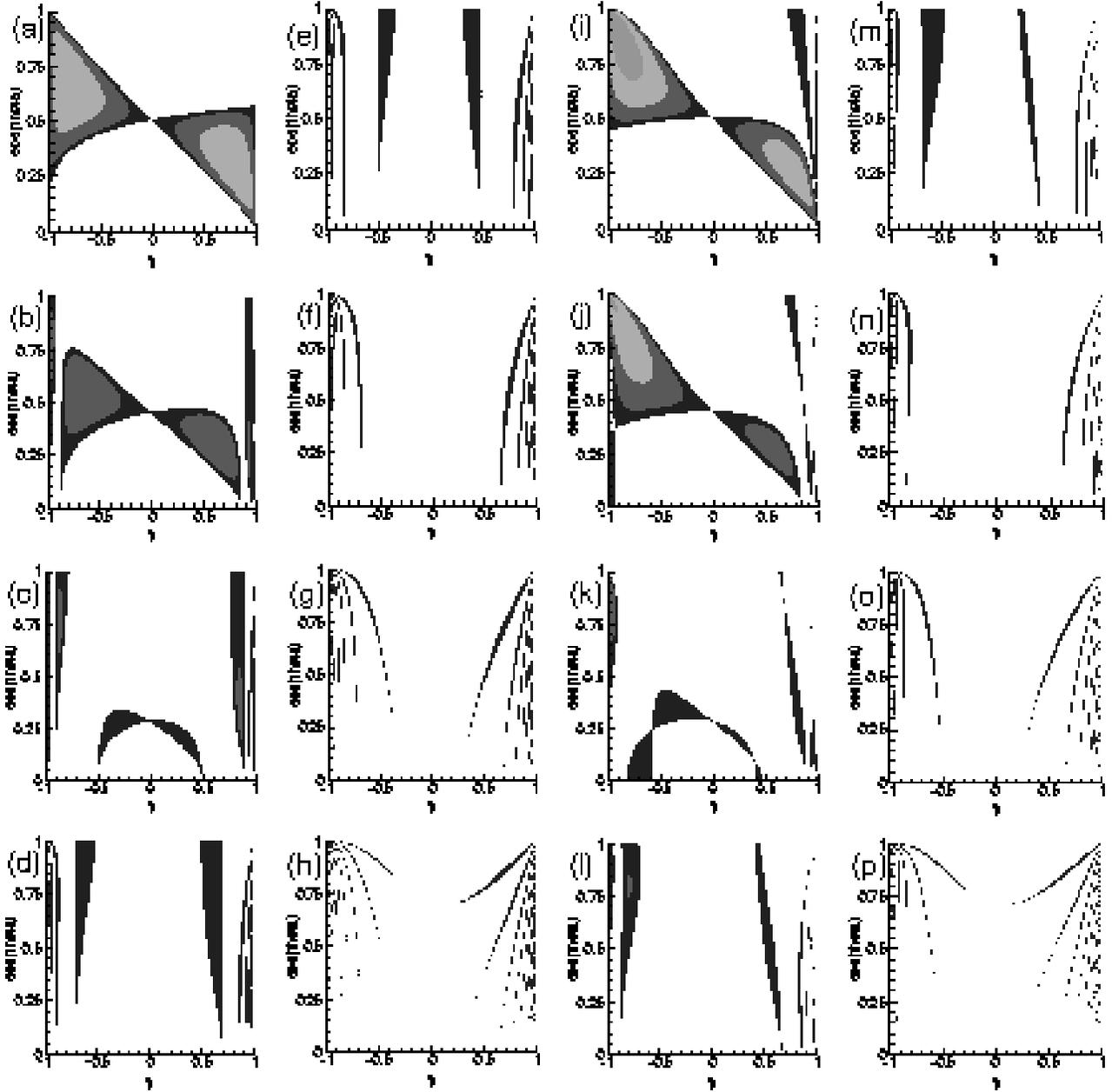}
\caption{Instability domain of $\avec(t)$ in  
   the classical EB equations (Figs.(a)-(h) with $\Upsilon_0 = 1$) and
   the \laeba~model (Figs. (i)-(p) with $\Upsilon_0 = 5/4$)
   for rotation parameter $\zeta = 2$ (no rotation)
   and various values of buoyancy parameter $\chi$:
   (a),(i)~0; (b),(j)~1/4; (c),(k)~3/4; (d),(l)~2;
   (e),(m)~3; (f),(n)~5; (g),(o)~8; and (h),(p)~14.
   Regions in white represent parameter values for which the amplitude
   $\avec$ is bounded in
   time. For all other regions, the amplitude grows exponentially in
   time.  The contour lines are level lines of the Lyapunov-like growth
   rate with the same coloring in each figure.  The fact that some of
  the fingers appear to be disjoint regions is merely an artifact of
  numerical simulation and data visualization.   
  See the text for a description of the behavior of each individual
   subfigure.   The values of $N^2$ and $\Omega$ in (a)-(h) differ
   from those in (i)-(p) due to the fact that $\chi$ and $\zeta$ are
   similarity variables (see \eqref{eq:newdefs}).  
\label{fig:baylyN2-EB}
}
\end{center}
\end{figure*}

\begin{figure*}
\begin{center}
\includegraphics[width=6.8in]{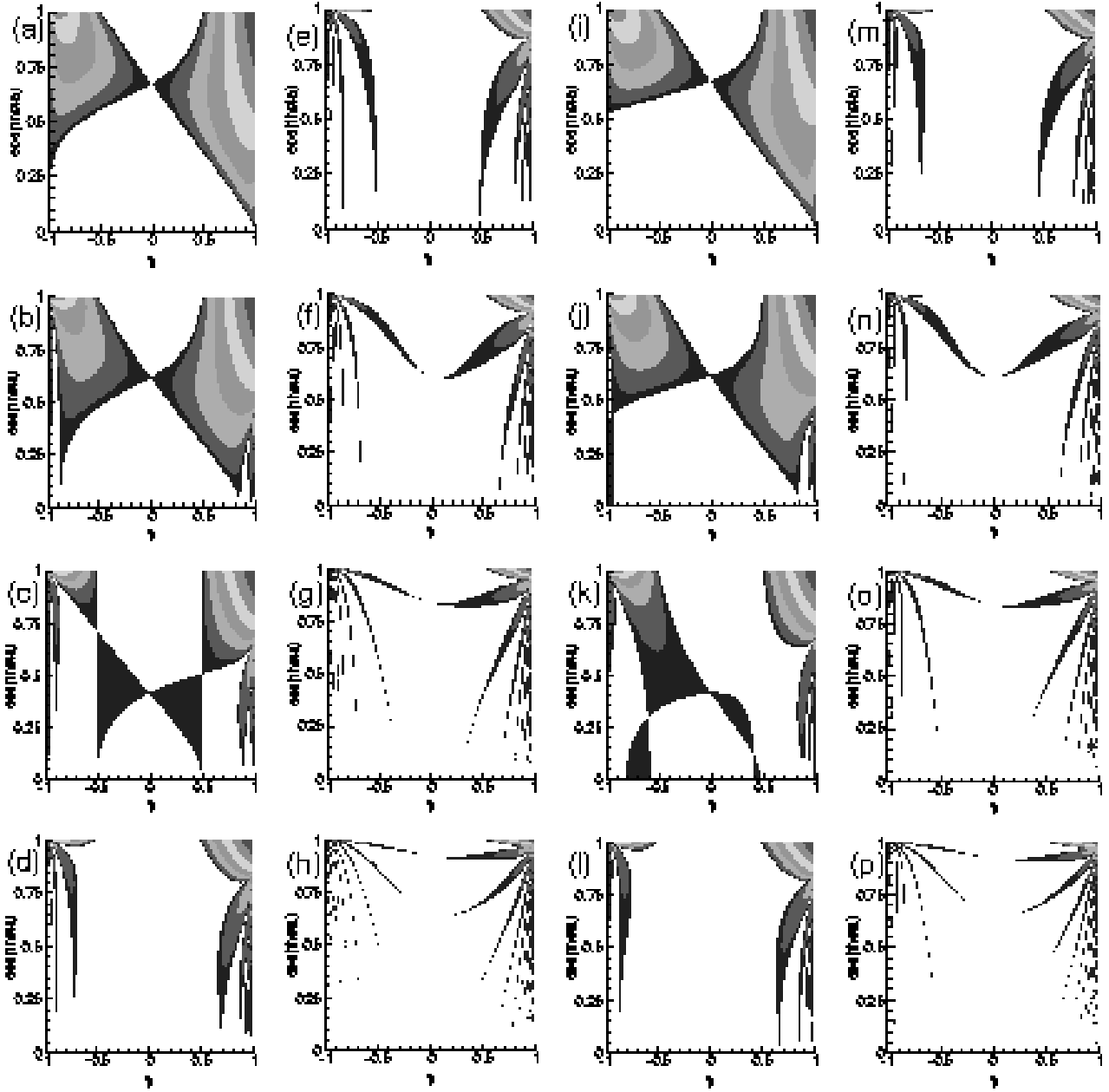}
\caption{The instability domains of $\avec(t)$ in  
   the classical EB equations (Figs.(a)-(h) with $\Upsilon_0 = 1$) and
   the \laeba~model (Figs.(i)-(p) with $\Upsilon_0 = 5/4$) for
   rotation parameter $\zeta = 3/2$
  and the same values of buoyancy parameter 
  $\chi$ as in Fig.~\ref{fig:baylyN2-EB}:
   (a),(i)~0; (b),(j)~1/4; (c),(k)~3/4; (d),(l)~2;
   (e),(m)~3; (f),(n)~5; (g),(o)~8; and (h),(p)~14.
  Note the presence of a stable band in (d)-(e) and (l)-(m), corresponding to 
  $1 < \sqrt\chi < 2$ (see \eqref{eq:stabcrita}). 
\label{fig:baylyN2z15-EB}
}
\end{center}
\end{figure*}

\begin{figure*}
\begin{center}
\includegraphics[width=6.8in]{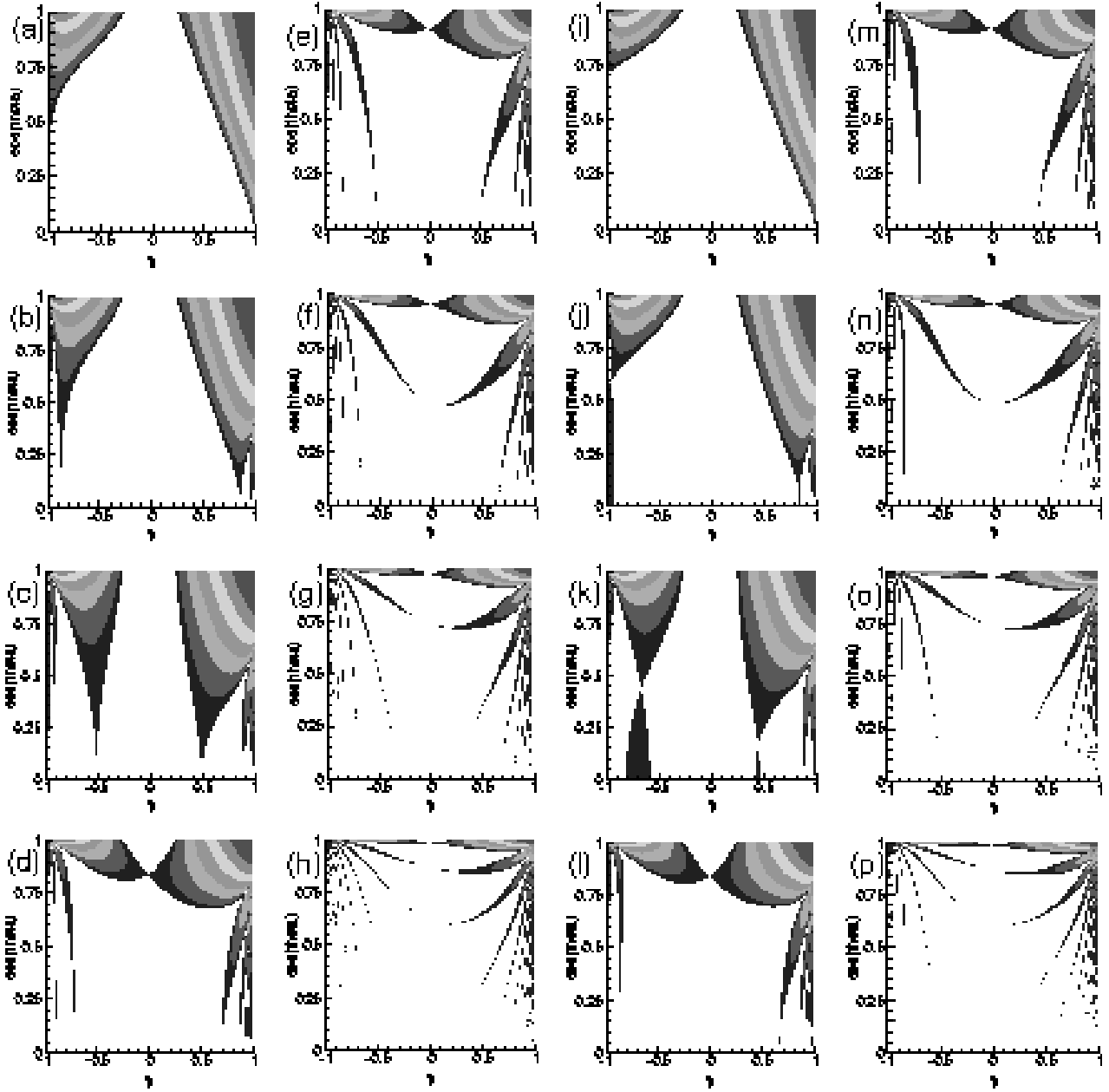}
\caption{The instability domains of $\avec(t)$ in  
   the classical EB equations (Figs.(a)-(h) with $\Upsilon_0 = 1$) and
   the \laeba~model (Figs.(i)-(p) with $\Upsilon_0 = 5/4$) for
   rotation parameter $\zeta = 3/4$
  and the same values of buoyancy parameter 
  $\chi$ as in Fig.~\ref{fig:baylyN2-EB}:
   (a),(i)~0; (b),(j)~1/4; (c),(k)~3/4; (d),(l)~2;
   (e),(m)~3; (f),(n)~5; (g),(o)~8; and (h),(p)~14.
  Note that since $\zeta < 1$, we have a stable band of flows in
  (a)-(c) and (i)-(k), where $\chi < 1$ (see \eqref{eq:stabcritb}). 
\label{fig:baylyN2z075-EB}
}
\end{center}
\end{figure*}

Figures~\ref{fig:baylyN2-EB}-\ref{fig:baylyN2z075-EB} display the
instability domains for a representative set of $\zeta$ and $\chi$.
We reiterate that subfigures (a)-(h) correspond to the classic EB
equations ($\Upsilon_0 = 1$), and subfigures (i)-(p) correspond to the
\laeba~model for parameter value $\Upsilon_0 = 5/4$. 
\newt{The figures are set side-by-side so that the reader can easily
  see the effects the model has on the classical solutions.}   
We see the large number of instability fingers, or Arnold
tongues, corresponding to various values of $n$ in
\eqref{eq:critangle}. The previously described shift of fingers  into 
and out of regions (I)-(IV) is also clearly seen.

We describe the behavior of the fingers in  
Fig.~\ref{fig:baylyN2-EB}, the case $\zeta =2$ ($\Omega = 0$), in detail.  
For $\chi=0$ (Fig.~\ref{fig:baylyN2-EB}a,i), the
  critical angles are $\cos\theta = 1/2$ (subharmonic) and
  $\cos\theta = 1$ (fundamental), the latter being true for all $\chi$. 
  The subharmonic finger lives in region
  (I), and the fundamental finger lives on the boundary between
  regions (I) and (II).  As $\chi$ increases towards unity
  (Figs.~\ref{fig:baylyN2-EB}b,c,j,k), the
  subharmonic finger shifts in the parameter plane towards $\cos\theta
  = 0$. 
  In the range $1 < \chi < 4$
  (Figs.~\ref{fig:baylyN2-EB}d,e,l,m), the
  finger corresponding to the 
  subharmonic frequency has vanished, that
  is, it has entered 
  region (IV).  We conjecture that the vertically aligned finger in
  these pictures, which collapses on the line $\gamma = 0$ at $\chi =
  4$, corresponds to the fundamental frequency ($n=2$).  In the range
  $4 <\chi < 9$ (Figs.~\ref{fig:baylyN2-EB}f,g,n,o), the fingers corresponding to the
   superharmonic frequencies ($n \geq 3$)
  have shifted from region 
  (II) to region (IV) and appear to be vertical.  For $\chi > 9$
  (Figs.~\ref{fig:baylyN2-EB}h,p), 
  the fingers corresponding to the first superharmonic frequency ($n= 3$)
  have shifted from region 
  (IV) to region (I) and meet at the prescribed critical angle
  \eqref{eq:critangle} along
  $\gamma = 0$.  Once
  again, the flow is unstable for infinitesimally small, non-zero 
  eccentricities.  Seen
  is the second superharmonic ($n=4$) shifting in the plane.
  What cannot be seen is 
  the fact that the finger for the subharmonic 
  frequency ($n=1$) has disappeared into region (III), never to return.
  Not shown is the fact that the second superharmonic frequency's
  fingers ($n=4$) will touch on $\gamma = 0$ 
   at $\chi = 16$.  Similarly for the third superharmonic at $\chi
  = 25$, and so on.  The fixed
  value of $\zeta = 2$ satisfies \eqref{eq:stabcrit3} for all
  $\gamma$, and thus 
  the flow is always stable on the line $\cos\theta = 1$.  The
  vertical lines in Fig.~\ref{fig:baylyN2-EB}b-e,i-m, corresponding to
  the second finger, do not touch the line $\cos\theta =1$.  Rather,
  they curve sharply near that line, meeting it only at $\gamma=1$.  
  This behavior is difficult to resolve numerically.  However, we
  verify that this case is stable on the line $\cos\theta = 1$
  itself.  
The behavior of the fingers is similar in the presence of rotation
(Figs.~\ref{fig:baylyN2z15-EB} and \ref{fig:baylyN2z075-EB}).  

Figure~\ref{fig:baylyN2z15-EB} shows corresponding contour plots for the
parameter value $\zeta =3/2$.  We see that
the primary finger (subharmonic frequency)  touches the $\gamma=0$
line at $\cos\theta = 2/3$ 
(Figs.~\ref{fig:baylyN2z15-EB}a,i).
Note that
for $1 < \chi < 4$ (Figs.~\ref{fig:baylyN2z15-EB}d,e,l,m), there are eccentric flows  for which
the Kelvin waves are stable.  This is because 
\eqref{eq:stabcrita} is satisfied for these parameter values.  
The value of $\zeta$ here satisfies
\eqref{eq:stabcrit3} only for $-1/2<\gamma<1/2$.  Outside of this
region, the flow is unstable on the line $\cos\theta = 1$.  We see
that in all images, this fact holds true.  Any visual deviation from this
stability band is a consequence of the plotting program.  
We verify
numerically that the band does not change width as a function of
$\Upsilon_0$ or $\chi$.  

The behavior for $\zeta = 3/4$, shown
in Fig.~\ref{fig:baylyN2z075-EB}, 
is similar to the previous cases.  Note that the main difference here
is the stable band of eccentric flows for $\chi < 1$
(Figs.~\ref{fig:baylyN2z075-EB}a-c,i-k), where  \eqref{eq:stabcritb} is 
satisfied.  Furthermore, we see pairs of  fingers enter
region (I) as predicted--the primary (subharmonic) 
for $\chi > 1$, the secondary (fundamental) for
$\chi > 4$, and the tertiary (first superharmonic) 
for $\chi > 9$ (Figs.~\ref{fig:baylyN2-EB}d-h,l-p).

{\it Remark.} Previous work\cite{fab:holm:04a} commented on the fact that there
is a remarkable similarity between the effects of counter-rotation,
i.e. decreasing $|\Omega+1|$, and
increasing the parameter $\alpha$  on elliptic
instability.  
This similarity is better understood by choosing $\zeta =
2|\Omega+1|/(1+\alpha^2\beta^2)$ as a similarity variable.   
Numerical
simulations show that the growth rates obtained by maximizing over the
$(\gamma,\cos\theta)$-plane are roughly constant functions of $\Upsilon_0$ for
randomly chosen fixed values of $\chi$ and $\zeta$ satisfying $\zeta <
1$.  This is because the maximum typically occurs on the finger 
associated with the subharmonic frequency ($n=1$).    
Again, this indicates that the \laeba~model will preserve
the basic characteristics of 
elliptic instability  by properly adjusting the Rossby number and the
Brunt-V\"ais\"al\"a frequency.  When $\zeta > 1$, however, the finger
associated with the subharmonic frequency has vanished, and the
maximum growth rates vary due to the presence of other frequencies.
Since the model alters the shape of the fingers associated with these
frequencies as well as the associated Lyapunov exponents for moderate
and large eccentricities, there is no reason to expect that the growth
rate remain constant.  

\newt{%
\section{Anisotropic \laeba~model}
We now examine the effects of anisotropy in the third component of the
velocity field $\vvec$.  We do this by altering the definition of the
transported velocity $\vvec$ from $\vvec = (1-\alpha^2\Delta)\uvec$ to  
\begin{eqnarray}
\vvec = (1 - \alpha^2(\partial^2_{xx} + \partial^2_{yy} +
\epsilon^2\partial^2_{zz}))\uvec.  
\end{eqnarray}
Here, $\epsilon$ measures the anisotropy of the transported velocity
field $0 \leq \epsilon \leq 1$.  For $\epsilon = 1$, we regain the
results of the previous section.  From the view of elliptic
instability, the main difference is the change in the definition of
$\Upsilon(t)$ from \eqref{eq:upsilondef} to 
\begin{eqnarray}
\Upsilon(t) = 1+\alpha^2\beta^2(k_1^2(t) + k_2^2(t) +
\epsilon^2k_3^2(t)).  
\end{eqnarray}
The theory for the isotropic \laeba~model
in the parameter regime $|\gamma|\ll 1$ 
remains intact using the similarity variables given in
\eqref{eq:newdefs} with 
\begin{eqnarray}\label{eq:newup0}
\Upsilon_0 = 1+\alpha^2\beta^2(1-(1-\epsilon^2)\cos^2\theta).  
\end{eqnarray}
Effectively, anisotropy in the third component will reduce the value
of $\Upsilon_0$ (since $1-(1-\epsilon^2)\cos^2\theta\leq 1$).
Numerical simulations verify that there are no changes along the line
$\gamma=0$, trival differences in the region
$|\gamma|\ll 1$, and possibly noticable differences for $\gamma
\approx 1$, the last of which 
occur only for values $\alpha^2\beta^2> 1$.  The main distinction
lies in the fact that each subfigure of 
Figs.~\ref{fig:baylyN2-EB}-\ref{fig:baylyN2z075-EB} corresponds to 
fixed values of $\Omega$ and $N^2$, whereas similar figures using
\eqref{eq:newdefs} with \eqref{eq:newup0} would result in different
horizontal lines, that is, each value of $\cos\theta$,  
corresponding to different $\Omega$ and $N^2$ values within a
subfigure. 
}

\section{Conclusions}
We have studied the effects of a turbulence closure model for 
the Euler-Boussinesq equations on elliptic instability.  We found a
trade-off in the effects of turbulence, in that
the model can increase the width of one stability band (see
\eqref{eq:stabcrita}) while simultaneously decreasing the width of
another (see \eqref{eq:stabcritb}).  
This occurs because of wave number dependence is introduced when  the
smoothing operation (with lengthscale $\alpha$) is applied in developing
a mean model of turbulence. We expect that other turbulence models
will show similar tendencies, although our earlier work \cite{fab:holm:04a}
shows that the magnitudes and functional forms of these tendencies may
vary from one turbulence model to another. 

The \laeba~model
considered here has the advantage of preserving the main features of
classic elliptic instability analysis for circular flows, 
up to rescaling the stability
conditions in terms of rotation and buoyancy frequencies by certain
factors involving the nondimensional wavenumber parameter
$\Upsilon_0$, \newt{which may also contain anisotropic information as
  well.}  
This wavenumber parameter is made nondimensional by using
the lengthscale (turbulence correlation width) in the \laeba~model. The
similarity property and the preservation of structure under
appropriate rescaling the frequencies by $\Upsilon_0$ has
allowed a complete analysis of the mean effects of turbulence on the
elliptic instability, when viewed as an exact nonlinear CC solution of
the \laeba~model. This similarity property also reveals how the
smoothing introduced in developing the turbulence model
interacts with the physical effects of rotation and buoyancy, which
are vital for understanding turbulence in geophysical flows. One
outstanding question remains, which is something of a technical
detail. Namely, we would still like to know the mechanism by which the
turbulence model affects the resonance regions seen
in Figs.~\ref{fig:baylyalpha1}-\ref{fig:baylyalpha2}  for elliptical
flows of high 
eccenticity.  
These are similarity breaking effects of the \laeba~turbulence
model. They arise because of the introduction of the length
scale $\alpha$ and cannot be removed by rescaling the other
parameters in the problem. \newt{Anisotropy in the $z-$component of the
velocity field amplifies these effects.}  
These effects tend to occur as the shear in
the flow (represented by $\gamma$) becomes more pronounced.  
In flows of such high eccenticity, coursening at the correlation
length $\alpha$ in the turbulence model may be exhibiting different
effects than for small eccentricity, because of the higher anisotropy
of these flows \newt{in the plane of the strain flow}.  
The derivation of the \laeba~model assumes
isotropy of the fluctuations \newt{in all
  coordinates}, which does not hold in the case of flows
with high eccentricity.  Perhaps another version of the model--one
which would allow for strong \newt{planar} anisotropy in the statistical
correlations of turbulent Lagrangian fluctuations with their mean
trajectories--would not show such pronounced effects at high
eccentricity.  

Additionally, the \laeba~model introduces a band of unstable flows in the
parameter regime $\cos\theta = 0$, where the wave vector lies in the
plane of the flow.  This is a direct consequence of the presence of
buoyancy.  Classically, this parameter regime is critically stable.  

\newt{Finally, we note that 
Miyazaki and Adachi\cite{miya:ada:98} were able to extend the result
of Lifschitz and Fabijonas\cite{lif:fab:96} to stratified circular
flows.  Namely, the stable parameter values for the Kelvin wave in the
circular case ($\gamma = 0$) are unstable with respect to
high-frequency perturbations.  We conjecture that the same holds true
for the \laeba~model. } 

\acknowledgments
We thank A. Lifschitz-Lipton for sparking our original interest 
in CC-solutions, Y. Fukumoto for his help in deriving the coupled
pair of Schr\"odinger equations, \newt{and the anonymous referees whose
comments help improve the paper, in particular the inclusion of
Section VI}.
This work was carried out while one of the authors
(BRF) visited the Theoretical Division of the 
Los Alamos National Laboratory.  We acknowledge funding from the
Division's Turbulence Working Group. DDH is grateful for support by US
DOE, under contract number 
W-7405-END-36 for Los Alamos National Laboraory, and Office of Science
ASCAR/AMS/MICS.  

{\it Numerics.}  The system of different equations was solved
numerically using the {\sc LSODE} solver \cite{dvode}, and eigenvalues
of the monodromy matrix were computed using {\sc LAPACK}
\cite{lapack}. We are grateful to the Center for Scientific Computing
at Southern Methodist University for use of their facilities.    

\appendix
\section{Recasting Eq.~(20) as a pair of Sch\"odinger
  equations}
We recast \eqref{eq:system} as a coupled pair of Sch\"odinger
  equations in the spirit of Ref.~\onlinecite{bay:holm:lif:96} for all base flows
  for which the matrix $\Smat$ has the form  
\begin{eqnarray*}
\Smat = \begin{pmatrix} \Lmatp & \zerovec_2 \\ \zerovec_2^T & 0 \end{pmatrix}.
\end{eqnarray*}
Here $\zerovec_2$ is the zero vector in $\mathbb{R}^2$ and $\Lmatp$
is a $2\times 2$ matrix.  Incompressibility demands that
$\operatorname{tr}(\Lmatp) = 0$.  

We decompose the amplitude vector $\avec$ and the wave vector $\kvec$
into components which are perpendicular and parallel to the axis of
rotation:  
\begin{eqnarray*}
\kvec = \begin{pmatrix} \kvecp \\ \kpll \end{pmatrix}, \qquad
\avec = \begin{pmatrix} \avecp \\ \apll \end{pmatrix}, \qquad
\Ovec = \begin{pmatrix} \zerovec_2 \\ \Omega \end{pmatrix},
\end{eqnarray*}
where $\zerovec_2$, $\kvecp$ and $\avecp$ are vectors in $\mathbb{R}^2$.
Equations~\eqref{eq:incompcond} and \eqref{eq:kevolg}-\eqref{eq:cevolg1}
take the form  
\begin{eqnarray*}
&&\kvecp^T\cdot\avecp + \kpll\apll = 0, \\
&&d_t \kvecp = -\Lmatp^T\cdot\kvecp, \qquad d_t\kpll = 0, \\
&&d_t (\Upsilon\avecp) = \frac{N^2\kpll\kvecp}{|\kvec|^2}\,\hat b \\
&&\phantom{X} 
 -\Big ( (\Upsilon-1)\Lmatp^T + \Lmatp 
 + 2\Omega\Rmat - \frac{\kvecp\kvecp^T\cdot\Mmat}{|\kvec|^2}\Big )
      \cdot  \avecp  , \\
&&d_t(\Upsilon\apll) =  
  \frac{\kpll\kvecp^T\cdot\Mmat}{|\kvec|^2}\cdot \avecp + 
  \Big ( -N^2 + \frac{N^2\kpll^2}{|\kvec|^2}\Big )\, \hat b, \\
&&d_t\hat b = \apll.
\end{eqnarray*}
Here, 
\begin{eqnarray*} 
&\Mmat = (\Upsilon+1)\Lmatp + (\Upsilon-1)\Lmatp^T + 2\Omega\Rmat, \\
&\Rmat = \begin{pmatrix} 0 & -1 \\ 1 & 0 \end{pmatrix}.\label{eq:rmat}
\end{eqnarray*}

Consider the transformations
\begin{eqnarray*}
&&p = 
\frac{|\kvec|}{|\kvecp|}\kvecp^T\cdot(\Upsilon\avecp) =
-\frac{|\kvec|}{|\kvecp|}\kpll(\Upsilon\apll), \label{eq:pdef}
\\
&&q = \frac{|\kvec|}{|\kvecp|}\left ( \kvec\times(\Upsilon\avec) \right
)_\parallel, \qquad r = \frac{|\kvec|}{|\kvecp|}\, \hat b.
\end{eqnarray*}
Here, $(\uvec)_\parallel$ represents the third component of a vector
$\uvec$.  The system of equations satisfied by these variables is 
\begin{eqnarray}\label{eq:transsys}
\frac{d}{dt}
\begin{pmatrix} p \\  q \\ r \end{pmatrix} = 
\begin{pmatrix} K & H & M_1 \\ -Q & -K & 0 \\ -M_2 & 0 & -K
\end{pmatrix} 
\begin{pmatrix} p \\   q \\ r
\end{pmatrix}.
\end{eqnarray}
where 
\begin{align*}
K &=  \frac{d}{dt} \left ( \ln \Big (
    \frac{|\kvecp|}{|\kvec|}\Big )\right ), \quad
H =
    \frac{\kpll^2\kvecp^T\cdot\Mmat\cdot\Rmat^T\cdot\kvecp}{\Upsilon|\kvec|^2|\kvecp|^2},\\
Q &= \frac{1}{\Upsilon}(W+2\Omega), \quad W =
    (\operatorname{curl}\,(\Lmat\cdot\xvec))_\parallel = (L_{21} -
    L_{12}) \\
M_1 &= \frac{\kpll N^2|\kvecp|^2}{|\kvec|^2}, \quad
M_2 = \frac{1}{\kpll\Upsilon}.
\end{align*}
We eliminate the variable $p$ from the problem,
recasting the above system as a pair of second order Schr\"odinger
equations for $q$  and $r$ 
(overhead dot denotes differentiation with respect to time):
\begin{eqnarray*}
\ddot q  - \frac{\dot Q}{Q}\dot q + \Big [ QH + \dot K - K^2 -
  \frac{\dot Q}{Q}K\Big ]\,q  + QM_1 \, r = 0 \\
\begin{split}
\ddot r + K\dot r + [\dot K + M_1M_2]\, r 
  - \frac{\dot M_2 + M_2 K}{Q}\,\dot q  \phantom{+ XXX}\\
  - \Big [ \frac{\dot M_2 K + M_2K^2}{Q} + M_2H \Big ]\, q = 0 
\end{split}
\end{eqnarray*}
The results for the \lansa~model are obtained by setting $N^2=0$ 
(corresponding to $M_1 = 0$) and ignoring the equation for $r$.  
The results for the Navier-Stokes equations\cite{bay:holm:lif:96}
are regained by setting $\Upsilon = 1$.

\clearpage

\begin{figure}
\begin{center}
\includegraphics[width=3.4in]{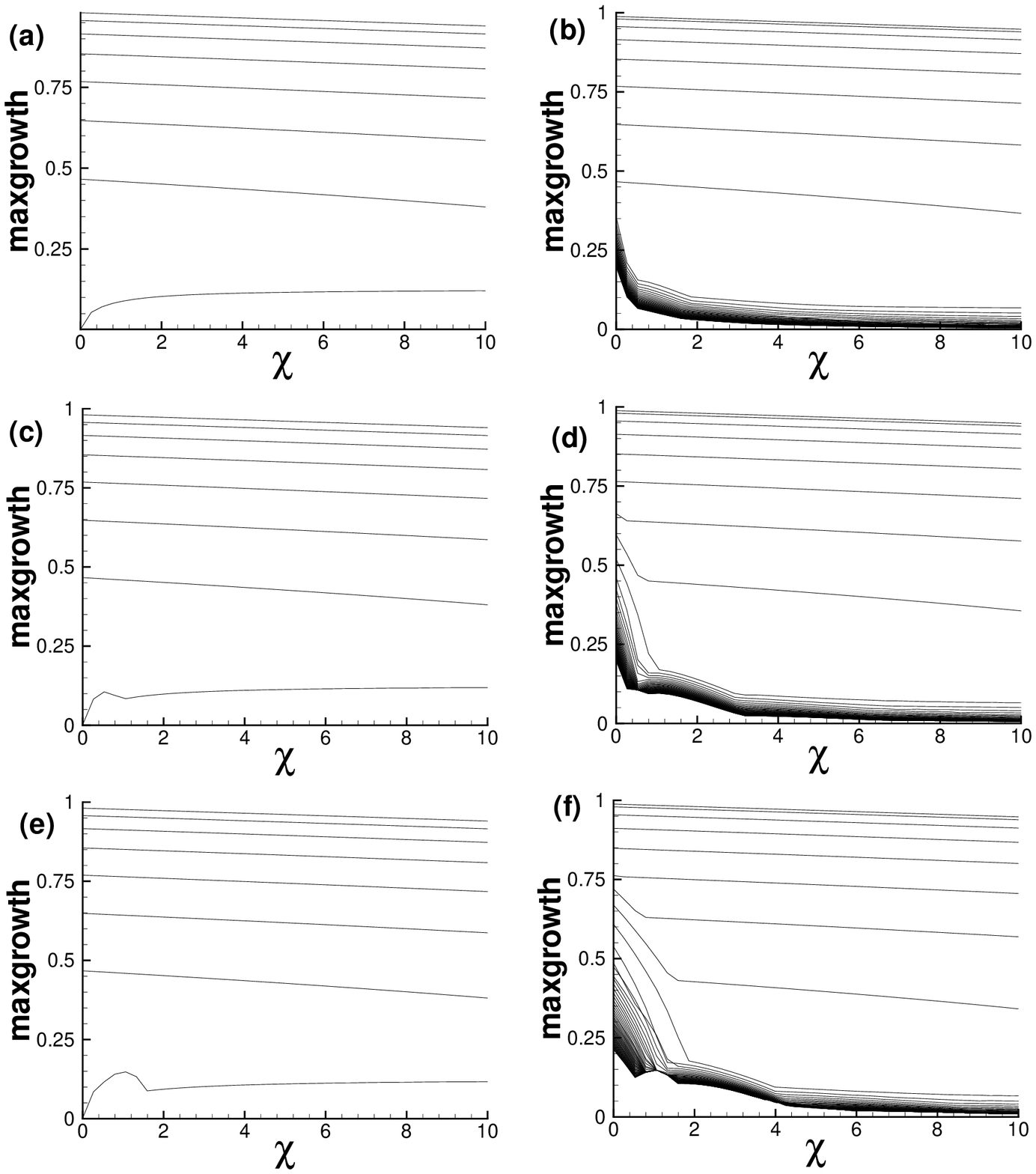}
\caption{The growth rate maximzed over the $(\cos\theta,\gamma)$ plane
as a function of $\chi$ for various
$\zeta$ and $\Upsilon_0$: (a-b) $\Upsilon_0=1$, (c-d)
$\Upsilon_0=3/2$, (e-f) $\Upsilon_0=4$.   
The figures in the left column 
correspond to $\zeta$ values satisfying $0 \leq \zeta \leq 1$, 
increasing from bottom to top. 
Similarly, the figures in the right column 
correspond to $\zeta$ values satisfying $\zeta > 1$,
increasing from top to bottom.  Notice that as
$\zeta\to\infty$, the growth rates approach an asymptotic limit for
all $\Upsilon_0$.  This
is in accordance with the Taylor-Proudman theorem. 
Notice further the development of a ``bump'' for $\Upsilon_0 > 1$, when
$\zeta = 0$ (the bottom-most curve in the figures in the left column) 
and as $\zeta \to \infty$ (the bottom-most curves in the figures in
the right column).  
\label{fig:sigmamaxN2} }
\end{center}
\end{figure}


\begin{thebibliography}{10}

\bibitem{fab:holm:03}
B.~R. Fabijonas and D.~D. Holm,
\newblock ``Mean effects of turbulence on elliptic instability,''
\newblock Phys. Rev. Lett. {\bf 90}, 124501 (2003).

\bibitem{fab:holm:04a}
B.~R. Fabijonas and D.~D. Holm,
\newblock ``Craik-{C}riminale solutions and elliptic instability in
  nonlinear-reactive closure models for turbulence,''
\newblock Phys. Fluids {\bf 16}, 853 (2004).

\bibitem{holm:marsden:rat:98a}
D.~D. Holm, J.~E. Marsden, and T.~S. Ratiu,
\newblock ``Euler-{P}oincar\'e models of ideal fluids with nonlinear
  dispersion,''
\newblock Phys. Rev. Lett. {\bf 80}, 4173 (1998).

\bibitem{holm:99}
D.~D. Holm,
\newblock ``Fluctuation effects on 3d lagrangian mean and eulerian mean fluid
  motion,''
\newblock Physica {\bf 133D}, 215 (1999).

\bibitem{holm:mar:rat:02}
D.~D. Holm, J.~E. Marsden, and T.~S. Ratiu,
\newblock ``The {E}uler-{P}oincar\'e equations in geophysical fluid
  dynamics,''
\newblock in {\em Large-Scale Atmosphere-Ocean Dynamics 2: Geometric Methods
  and Models}, edited by J.~Norbury and I.~Roulstone, pages 251--299, Cambridge
  University Press, 2002.

\bibitem{leray:34}
J.~Leray,
\newblock ``Sur le mouvement d'un liquide visqueux emplissant l'espace,''
\newblock Acta Math. {\bf 63}, 193 (1934).

\bibitem{kelvin}
Lord~Kelvin,
\newblock ``Stability of fluid motion: rectilinear motion of viscous fluid
  between two parallel plates,''
\newblock Phil. Mag. {\bf 24}, 188 (1887).

\bibitem{bayly:86}
B.~J. Bayly,
\newblock ``Three-dimensional instability of elliptical flow,''
\newblock Phys. Rev. Lett. {\bf 57}, 2160 (1986).

\bibitem{craik:crim:86}
A.~D.~D. Craik and W.~O. Criminale,
\newblock ``Evolution of wavelike disturbances in shear flows: a class of exact
  solutions of the {N}avier-{S}tokes equations,''
\newblock Proc. R. Soc. London A {\bf 406}, 13 (1986).

\bibitem{craik:88}
A.~D.~D. Craik,
\newblock ``A class of exact solutions in viscous incompressible
  magnetohydrodynamics,''
\newblock Proc. R. Soc. London A {\bf 417}, 235 (1988).

\bibitem{craik:89}
A.~D.~D. Craik,
\newblock ``The stability of unbounded two- and three-dimensional flows subject
  to body forces: some exact solutions,''
\newblock J. Fluid Mech. {\bf 198}, 275 (1989).

\bibitem{wal:90}
F.~Waleffe,
\newblock ``On the three-dimensional instability of strained vortices,''
\newblock Phys. Fluids A {\bf 2}, 76 (1990).

\bibitem{miya:fuku:92}
T.~Miyazaki and Y.~Fukumoto,
\newblock ``Three-dimensional instability of strained vortices in a stably
  stratified fluid,''
\newblock Phys. Fluids {\bf 4}, 2515 (1992).

\bibitem{kers:93}
R.~R. Kerswell,
\newblock ``Elliptical instabilities of stratified, hydromagnetic waves,''
\newblock Geophys. Astrophys. Fluid Dynam. {\bf 71}, 105 (1993).

\bibitem{miya:93}
T.~Miyazaki,
\newblock ``Elliptical instability in a stably stratified rotating fluid,''
\newblock Phys. Fluids {\bf 5}, 2702 (1993).

\bibitem{leb:03}
S.~Leblanc,
\newblock ``Internal wave resonances in strain flows,''
\newblock J. Fluid Mech. {\bf 477}, 259 (2003).

\bibitem{kers:02}
R.~R. Kerswell,
\newblock ``Elliptical instability,''
\newblock Annu. Rev. Fluid Mech. {\bf 34}, 83 (2002).

\bibitem{yaku:star:76}
V.~A. Yakubovich and V.~M. Starzhinskii,
\newblock {\em Linear Differential Equations with Periodic Coefficients},
\newblock Wiley, 1967.

\bibitem{cambon:94}
C. Cambon, J.~.P Beno\^it, L. Shao, and L. Jacquin,
\newblock ``Stability analysis and large eddy simulation of rotating
   turbulence with organized eddies,''
\newblock J. Fluid Mech. {\bf 278}, 175 (1994).

\bibitem{smith:wal:02}
L.~M. Smith and F.~Waleffe,
\newblock ``Generation of flow large scales in forced rotating stratified
  turbulence,''
\newblock J. Fluid Mech. {\bf 451}, 145 (2002).

\bibitem{miya:ada:98}
T.~Miyazaki and K.~Adachi,
\newblock ``Short-wavelength instabilities of waves in rotating
   stratified fluids,''
\newblock Phys. Fluids {\bf 10}, 3168 (1998).

\bibitem{lif:fab:96}
A.~Lifschitz and B.~R. Fabijonas,
\newblock ``A new class of instabilities of rotating fluids,''
\newblock Phys. Fluids {\bf 8}, 2239 (1996).

\bibitem{bay:holm:lif:96}
B.~J. Bayly, D.~D. Holm, and A.~Lifschitz,
\newblock ``Three-dimensional stability of elliptical vortex columns in
  external strain flows,''
\newblock Phil. Trans. R. Soc. Lond. A {\bf 354}, 1 (1996).

\bibitem{dvode}
P.~N. Brown, G.~D. Byrne and A.~C. Hindmarsh
\newblock ``{VODE}: A Variable Coefficient {ODE} Solver,''
\newblock SIAM J. Sci. Stat. Comput. {\bf 10}, 1038 (1989).

\bibitem{lapack}
E.~Anderson et al., 
\newblock {\em {LAPACK} Users Guide -- Release 3.0},
\newblock SIAM, 2000.

\end{thebibliography}
\end{document}